\documentclass[aps,twocolumn,tightenlines]{revtex4}
\usepackage[dvipdf]{graphicx}
\usepackage{color}
\usepackage{changepage}
\usepackage{amssymb,amsmath}
\usepackage{mathrsfs}
\usepackage{txfonts}

\begin{document}
\title{Residual amplitude modulation at the $10^{-7}$ level for ultra-stable lasers}
\author{Jonathan Gillot, Santerelli Falzon Tetsing-Talla, Séverine Denis, Gwenhaël Goavec-Merou, Jacques Millo, Clément Lacroûte, Yann Kersalé} 
\address{ FEMTO-ST, CNRS, Université de Bourgogne-Franche-Comté, ENSMM, 26, rue de l'Epitaphe, 25000 Besançon, France}

\date{\today}

\begin{abstract}
The stabilization of lasers on ultra-stable optical cavities by the Pound-Drever-Hall (PDH) technique is a widely used method. The PDH method relies on the phase-modulation of the laser, which is usually performed by an electro-optic modulator (EOM). When approaching the $10^{-16}$ level, this technology requires an active control of the residual amplitude modulation (RAM) generated by the EOM in order to bring the frequency stability of the laser down to the thermal noise limit of the ultra-stable cavity. In this article, we report on the development of an active system of RAM reduction based on a free space EOM, which is used to perform PDH-stabilization of a laser on a cryogenic silicon cavity.  A RAM stability of $1.4 \times 10^{-7}$ is obtained by employing a digital servo that stabilizes the EOM DC electric field, the crystal temperature and the laser power. Considering an ultra-stable cavity with a finesse of $2.5\times 10^5$, this RAM level would contribute to the fractional frequency instability at the level of about $5\times 10^{-19}$, well below the state of the art thermal noise limit of a few $ 10^{-17}$.
\end{abstract}
\maketitle
\bigskip

Lasers stabilized to ultra-stable optical cavities are widely spread devices for precise fundamental experiments like gravitational waves detectors \cite{Kokeyama2013,LIGO2020}, spectroscopy \cite{Ye98}, frequency standards  \cite{Young99,Stoehr06,Ludlow07,Alnis08,Kessler2012,Nicholson2012,Zhang2017,Wu2016} and tests of Lorentz invariance violation \cite{Muller2003b,Muller2003,Herrmann2009,Zhang2021}. The Pound-Drever-Hall (PDH) technique \cite{Drever1983} enables the frequency stabilization of lasers at a great level of precision, and the state of the art is better than $\sigma_{y}= 10^{-16}$ \cite{Kessler2012, Cole2013}.\\
\indent The continuous improvement of the stability of optical cavities faces several technical challenges \emph{e. g.} mechanical vibrations, fluctuations in laser power, or thermal noise. Among them, some stray effects grouped under the term of residual amplitude modulation (RAM) are responsible for an uncontrolled offset on the servo error signal that deteriorates the laser frequency stability. The PDH stabilization method requires phase modulation of the laser.
While recently demonstrated with an acousto-optic modulator (AOM) \cite{Zeng2021}, the phase modulation is generally applied by an electro-optic modulator crystal (EOM) which generates sidebands on the optical carrier. RAM arises from several origins \cite{Yu2016} including a polarization mismatch between the extraordinary axis of the EOM crystal and the polarization plane of the light \cite{Wong1985}, parasitic interferences in the EOM giving birth to etalon effects \cite{Whittaker1985, Duong2018}, etalon effects in optics downstream of the EOM \cite{Shen2015, Dangpeng2010} or some spatial inhomogeneities of the laser beam \cite{Sathian2013}.\\
\indent Passive suppression of RAM in EOMs has been demonstrated using wedged crystal ends \cite{Li2016, Tai2016, Bi2019, Jin2021}, and RAM can also be actively suppressed by acting on the EOM DC bias \cite{Wong1985, Li2014}, the EOM temperature \cite{Li2012}, or both \cite{Zhang2014}. Using digital electronics, we combine the stabilization of the laser power and EOM temperature to an active RAM suppression servo acting on the EOM DC input. We achieve a RAM level of $1.4 \times 10^{-7}$, compatible with a fracional frequency stability in the $10^{-19}$ range for a cavity finesse of $2.5\times 10^5$. 

\section{Analysis} \label{sec:analysis}

\indent The RAM in EOMs has been described theoretically in previous articles including \cite{Wong1985, Whittaker1985}, and several approaches and definitions can be found in the literature. 
We provide here the theoretical framework and definitions for the measurements presented in section \ref{sec:results}.

\begin{figure}[t!]
	\begin{adjustwidth}{0cm}{0cm}
		\center
		\includegraphics[width=8cm]{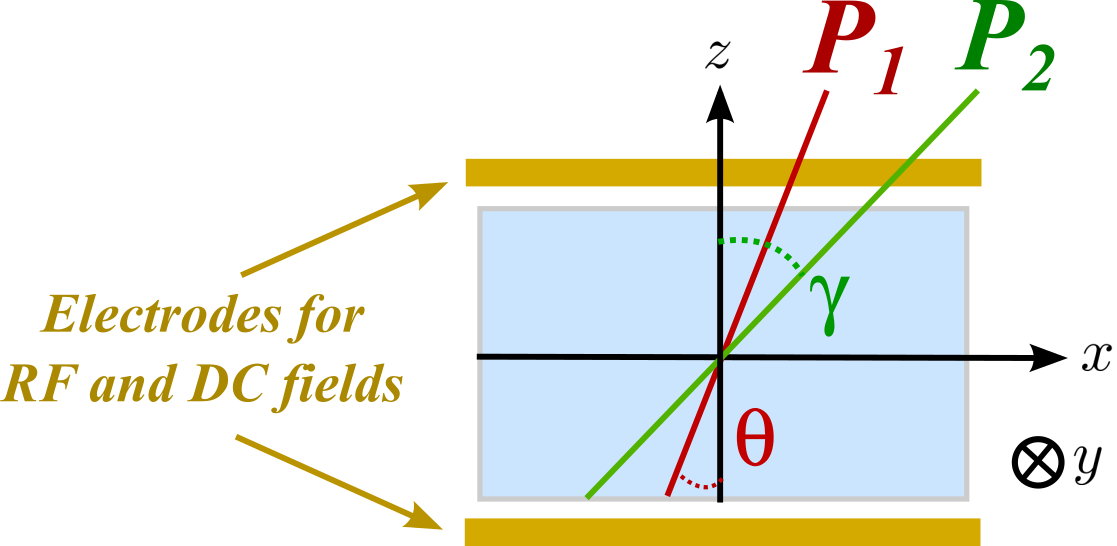}
	\end{adjustwidth}
	\caption{Schematic front view of the EOM crystal. Light propagates along the $y$-axis. The light field is modulated in a plane defined by the $z$-axis, parallel to the extraordinary axis of the birefringent crystal. If the input and output polarizers $P_1$ and $P_2$ are not perfectly aligned with the $z$-axis, a stray residual amplitude modulation depending on the $\theta$ and $\gamma$ angles appears.}
	\label{EOM}
\end{figure}

\indent The polarization of the laser beam before and after the EOM is defined with the help of two polarizers $P_1$ and $P_2$. The axis of $P_1$ and $P_2$ and the $z$-axis of the crystal are forming respectively the angles $\theta$ and $\gamma$ (Fig. \ref{EOM}). The expression of the optical field with amplitude $E_0$ projected onto the $z$-axis after the EOM output polarizer $P_2$  \cite{Wong1985} is:

\begin{equation} \label{equ-field}
    E(t) = E_0 e^{j \omega t} \left[ a e^{j (\Phi_x+\beta_x\sin(\Omega t))} + b e^{j (\Phi_z+\beta_z\sin(\Omega t))} \right]
\end{equation}

where $a = \sin\theta\sin\gamma$ and $b = \cos\theta\cos\gamma$ are the alignment factors sensitive to the temperature and stress-related effects, while $\omega$ and $\Omega$ are the angular frequency of the laser and the EOM radiofrequency (RF) modulation frequency, respectively. The phase shift due to the propagation in the crystal, that adds up to the shift induced by the applied DC voltage, is represented by $\Phi_x$ and $\Phi_z$ respectively for the $x$-axis and $z$-axis components. The phase shifts induced by the RF voltage are denoted $\beta_x$ and $\beta_z$. These 4 coefficients are related to the size of the crystal, electro-optic coefficients and the electric field applied across the crystal \cite{Wong1985}. Equation (\ref{equ-field}) can be developed to express the laser field:

\begin{equation} \label{equ-mod}
    E(t) = E_T \left( 1 + m_0e^{j\alpha} \sin(\Omega t) \right) e^{j (\omega t+ \beta_z \sin(\Omega t) )}.
\end{equation}

With the assumption that the polarizers are imperfectly aligned with the $z$-axis, $a \ll b \simeq 1$ and:
\begin{equation} \label{equ-mod-details}
   \left \{
   \begin{array}{r c l}
      m_0  & \simeq & 2\dfrac{a}{b} J_1(\xi)\\
      \alpha  & \simeq & \Delta\Phi+\pi/2\\
      E_T & \simeq & b E_0 e^{j\Phi_z}
   \end{array}
   \right .
\end{equation}

in which $J_1(\xi)$ is the first order Bessel function with $\xi=\beta_x-\beta_z$ the resulting modulation depth. $\Delta\Phi=\Phi_x-\Phi_z$ models all phase fluctuations induced by birefringence variations such that:

\begin{equation} \label{equ-phaseComp}
    \Delta\Phi = \Delta \phi_{\textrm{n}} + \Delta \phi_{\textrm{T}^\circ} + \Delta \phi_{\textrm{DC}}
\end{equation}
where $\Delta \phi_{\textrm{n}}$ is the natural birefringence, $\Delta \phi_{\textrm{T}^\circ}$ stems from the temperature variations and $\Delta \phi_{\textrm{DC}}$ from the DC voltage phase shift.

Equation (\ref{equ-mod}) describes an amplitude and phase modulated optical field, in which the amplitude modulation term $m=m_0e^{j\alpha}$ is complex and introduces a phase shift of the amplitude modulated signal. By measuring the field $E(t)$ described by Eq. (\ref{equ-mod}) with a photodiode of responsivity $\mathcal{R}$, we detect the RAM photo-current $i_{\textrm{RAM}}(t)$ at the modulation frequency $\Omega/2 \pi$:

\begin{equation} \label{eq-photocurrent}
    i_{\textrm{RAM}}(t)  = -4 \, \mathcal{R} \, a \, b \, K \, {E_0}^2 J_1(\xi) \sin(\Delta \Phi) \sin(\Omega t)
\end{equation}

with $K={J_0(\beta_z)}^2+2{J_1(\beta_z)}^2 \simeq 0.96$ for $\beta_z=1.08 \textrm{~rad}$, the optimum phase modulation depth for the PDH technique \cite{black2001}. $K$ differs from 1 because the development of the phase modulation term has been limited to the first order and is approximated to 1 in the following. Equation (\ref{eq-photocurrent}) shows that one can null the RAM induced photo-current by carefully aligning the axis of the polarizers with the axis of the crystal: $\gamma=0$ or $\theta=0$ leads to $a=0$ and $b=1$. However, even with an extreme precision in aligning the polarizations of the light with the axis of the crystal, fluctuations of temperature and mechanical vibrations are noticeably acting on the polarization factors $a$ and $b$. This fine alignment is thus degraded after a time that depends on the thermal insulation and the vibration attenuation of the optical table. Equation (\ref{eq-photocurrent}) shows that $i_{\textrm{RAM}}(t)$ is cancelled if $\Delta \Phi = 0$. A servo control acting on this phase by changing the EOM DC voltage or temperature is able to strongly reduce the RAM photo-current arising from polarization mismatch. Such control of $\Delta\Phi$ cancels only the real part of $m$:

\begin{equation}\label{eq-definition-M}
    M = \Re[m]\simeq -2 \dfrac{a}{b} \, J_1(\xi)\sin{\Delta\Phi}    
\end{equation}

and does not affect the modulus of the amplitude modulation index, $m_0$ in Eq. (\ref{equ-mod-details}). $M$ is identified as the effective amplitude modulation index that is used to characterize the RAM. The photo-current can be rewritten as follows:
$$ i(t) = I_0 + i_{\textrm{RF}}\sin{\Omega t}$$
with $I_0=\mathcal{R}P_{\lambda}$ the DC part of the signal and $P_{\lambda}$ the optical carrier power. By identification with Eq. (\ref{eq-definition-M}), the RF current peak value is $i_{\textrm{RF}}=2 \mathcal{R}P_{\lambda}M$. While the ratio $i_{\textrm{RF}}/I_0=2M$ is used in many publications to estimate the level of RAM, we use $M$ instead to avoid an overestimation by a factor 2. 


An error signal is obtained by applying $IQ$ demodulation to the signal provided by the photodiode and a low pass filter to reject components at $2\Omega$: 

\begin{equation}
   \left \{
   \begin{array}{r c l}
      V_{\textrm{I}}(t) & = & \frac{1}{2} \, A \, R \, k_{\textrm{m}} \, i_{\textrm{RF}} \cos{\varphi} \\
       & & \\
      V_{\textrm{Q}}(t) & = & -\frac{1}{2} \, A \, R \, k_{\textrm{m}} \, i_{\textrm{RF}} \sin{\varphi}
   \end{array}
   \right  .
\end{equation}

with $R$ the equivalent resistance loading the photodiode, $A$ the voltage gain of the RF amplification, $k_{\textrm{m}}$ the conversion factor of the $IQ$ mixer and $\varphi$ the phase of the demodulation signal. When this phase is close to 0, the in-phase voltage is equal to:
\begin{equation} \label{equ-error-signal}
    V_{\textrm{I}}(t) \simeq A \, R \, k_{\textrm{m}} \, \mathcal{R} \, P_{\lambda} M.
\end{equation}
$ V_{\rm{I}}$ can be used as an error signal: it cancels when $M=0$ and is linear around 0. Under this condition, the quadrature term becomes small and proportional to possible fluctuations of $\varphi$, which makes it less relevant.\\
\indent The impact of the RAM of the EOM when used in an ultra-stable laser based on a Fabry-Perot cavity can be estimated theoretically by looking at the contribution of $M$ to the error signal of the PDH lock. We used the method proposed in \cite{black2001} with an expression of the optical field corresponding to eqs. (\ref{equ-mod}) and (\ref{equ-mod-details}) to calculate the photo-current at frequency $\Omega/2\pi$ provided by the photodiode that collects the reflection of the cavity (we consider the case of a phase modulation frequency larger than the cavity linewidth and ${m_0}^2$ terms are neglected):
\begin{multline}\label{eq-errorSignal}
    i_{\textrm{PDH}} (\delta \nu) = -4 \mathcal{R} J_0(\beta_z) J_1(\beta_z) {E_0}^2 \sin({\Omega t}) \; \Bigg[ \Im[F(\delta \nu)] \\
    - M \; \bigg(\Re[F(\delta \nu)] \Big( \dfrac{1}{2\rho} + \rho \Big) - \dfrac{1}{2}\rho \bigg) \\
    + m_0 \, \cos({\Delta\Phi}) \; \Im[F(\delta \nu)] \; \Bigg( \dfrac{1}{2\rho} - \rho \Bigg) \Bigg]
\end{multline}

with $\rho=J_1(\beta_z)/J_0(\beta_z)$. $\Re[F(\delta \nu)]$ and $\Im[F(\delta\nu)]$ are respectively the real and imaginary part of the cavity coefficient of reflection $F(\delta\nu)$ where $\delta\nu$ is the frequency detuning between the laser and the cavity. After the demodulation, the first term of Eq. (\ref{eq-errorSignal}) is the PDH error signal used to lock the laser on the cavity. For small detunings with respect to the cavity linewidth $\Delta \nu_{\textrm{c}}$ and high finesses, we can linearize $\Im[F(\delta \nu)] \simeq \delta\nu / (\pi \, \Delta \nu_{\textrm{c}})$. The rest of the equation is the contribution of the RAM and can be split in two terms: one proportional to $M$ is cancelled by setting $\Delta\Phi=0$ thanks to the active feedback; the last one proportional to $m_0 \, \cos{\Delta\Phi} \; \Im[F(\delta\nu)]$ is cancelled by locking the laser on the cavity since $\Im[F(0)]=0$.
When the laser is locked to the cavity, the PDH error signal can therefore be expressed as the sum of the PDH and RAM error signals: 
\begin{eqnarray} \label{error}
\varepsilon (\delta \nu) &=& \varepsilon (\delta \nu)_{\textrm{PDH}}+\varepsilon (\delta \nu)_{\textrm{RAM}} \nonumber \\ 
\varepsilon (\delta \nu) &=& -4R_0\mathcal{R} k_{\textrm{m}} J_0(\beta_z) J_1(\beta_z) {E_0}^2\frac{\delta \nu}{\Delta \nu_c}\\\nonumber
&+& R_0\mathcal{R} k_{\textrm{m}} J_1(\beta_z)^2{E_0}^2 M \nonumber
\end{eqnarray}
The fractional frequency fluctuations can be expressed as:
\begin{eqnarray} \label{relat_tot}
\frac{\delta \nu}{\nu_c} &=& -\frac{\varepsilon (\delta \nu)}{ 4R_0\mathcal{R} k_{\textrm{m}} J_0(\beta_z) J_1(\beta_z) {E_0}^2} \frac{\Delta \nu_c}{\nu_c}\\
&+& \frac{1}{4}\frac{J_1(\beta_z)}{J_0(\beta_z)} \frac{\Delta \nu_c}{\nu_c} M \nonumber
\end{eqnarray}
with $\nu_c$ the laser absolute frequency and $\Delta \nu_c$ the full width at half maximum. The RAM contribution to the fractional frequency stability is thus:
\begin{equation} \label{sigmaram}
\sigma^{\textrm{RAM}}_y= \frac{1}{4}\frac{J_1(\beta_z)}{J_0(\beta_z)} \frac{\Delta \nu_c}{\nu_c}\sigma_{M}.
\end{equation}
This expression gives values inferior to the usual formula $\sigma^{RAM}_y=({\Delta \nu_c}/{\nu_c}) \times \sigma_{M}$ \cite{Zhang2014} in which a factor $J_1(\beta_z)/4J_0(\beta_z)\simeq 0.16$ is neglected.

\section{Experimental setup} \label{sec:expt}
\begin{figure}[h!]
	\begin{adjustwidth}{0cm}{0cm}
		\center
		\includegraphics[width=8.6cm]{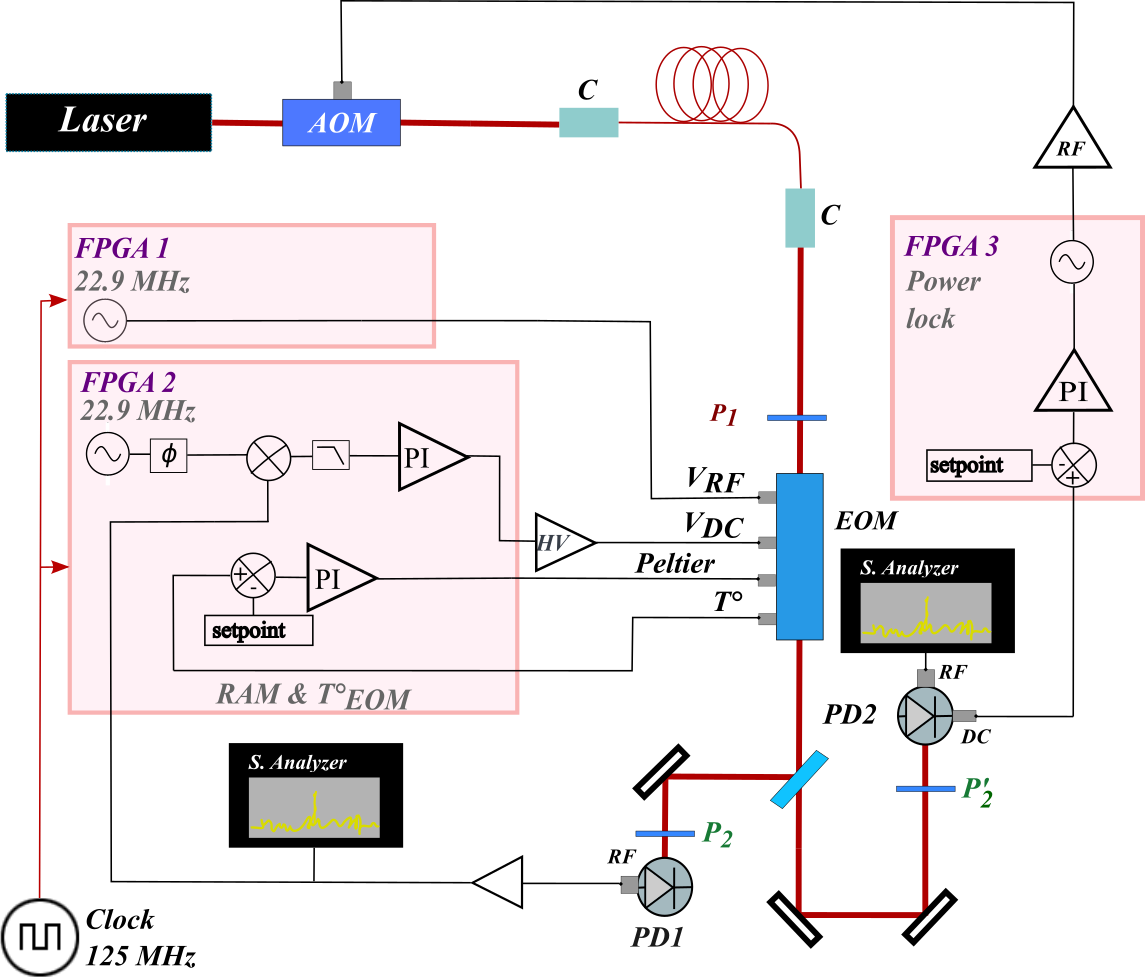}
	\end{adjustwidth}
	\caption{Experimental scheme for active RAM stabilization and characterization. C are collimators, HV is the high voltage generator, P$_1$, P$_2$ and P$_2'$ are polarizers, PD$1$ and PD$2$ respectively the in-loop and out-of-loop photodiodes.}
	\label{manip}
\end{figure}

Figure \ref{manip} depicts the experimental setup. 
We use a 1542~nm fiber laser. A free-space AOM placed at the output can be used to implement a power lock, as we know that fluctuating power has an incidence on the RAM since it can induce temperature effects in the EOM crystal and influence the stray etalon effect  \cite{Shen2015}. Power fluctuations are detected through the DC port of PD2 and corrected with the AOM RF power.

\indent After the output collimator $C$, a telescope resizes the beam and a power of $5$ mW is coupled to the free-space EOM. The polarization is finely tuned with a half-wave plate and a polarizer at the EOM input.
The phase modulation is provided by a LiNbO$_3$ EOM crystal on which we apply a RF power at $22.9$~MHz. The EOM is thermally controlled within $\pm 20$ mK with a Peltier device fastened below.
A 50/50 beam splitter sends half the light on PD1 for an active control of the RAM while the rest is sent on PD2 for an out-of-loop measurement. The beams are focused on the photodiodes with a waist estimated to be well below $1$ mm. The two photodiodes, with an active area diameter of 1~mm, are mounted on identical electronic cards. In order to test several configurations of the servo loop, a polarizer precedes each photodiode, instead of a unique polarizer placed at the output of the EOM. This configuration allows to test the response of the servo loop when the RAM signal is increased with the in-loop photodiode polarizer $P_2$ rotated by 45°.\\
\indent All along the optical path, optics are slightly tilted to an angle of $\sim 5 ^\circ$ to minimize parasistic etalon effects. The protective window of the in-loop photodiode has also been removed to eliminate a retro-reflection on the active area. Finally, in order to isolate the optical setup from air conditioning fluxes and dust, the cavity table is surrounded by an isolation box.

The out-of-loop RAM is directly measured and recorded with a spectrum analyzer  independently of the electronics used for the stabilization. In the in-loop branch, a $40 \textrm{~dB}$ amplifier and a directional $10 \textrm{~dB}$ coupler transmits the RF power $\mathcal{P}_m$ from photodiode PD1 to another spectrum analyzer for monitoring. 
Taking into account the load resistance of the photodiode $R_L$ and the input impedance $R_0$ of the RF amplifier, $M$ can be expressed as: 
\begin{equation} \label{M}
  M = \frac{R_L+R_0}{R_L \mathcal{R} A P_\lambda}\sqrt{\frac{ \mathcal{P}_m}{2R_0}}.
\end{equation}
The signal at $22.9 \textrm{~MHz}$ is sampled by an analog-to-digital converter (ADC, 14 bits, 125 MS/s) and transferred in a field programmable gate array (FPGA). The error signal of the control loop is obtained by carefully adjusting the phase of the demodulation. Synchronization between digital electronics boards is maintained by sharing the same clock signal at $125 \textrm{~MHz}$. The signal is then filtered and the data rate is reduced to $15.625 \textrm{~MS/s}$ before the proportional integrator function that produces the correction signal. The 14 bits digital to analog converter provides a $\pm 1 \textrm{~V}$ signal that is amplified by 46 dB and applied to the DC port of the EOM. With this large control voltage and a precise temperature regulation, the range of corrections is compatible with long time operation of the RAM control. We evaluate that the bandwidth of our RAM servo loop is close to $8$ kHz.\\
\indent The RAM lock and characterization has been integrated to our cryogenic cavity stabilized laser setup. With our $145$ mm long cavity, we expect that the thermal noise will limit the frequency stability at $3 \times 10^{-17}$ in fractional value \cite{Marechal2017}. This limit sets the goal to achieve for the RAM-induced fractional frequency instability.
 


\section{Results} \label{sec:results}

\begin{figure}[t!]
	\begin{adjustwidth}{0cm}{0cm}
		\center
		\includegraphics[width=8.3cm]{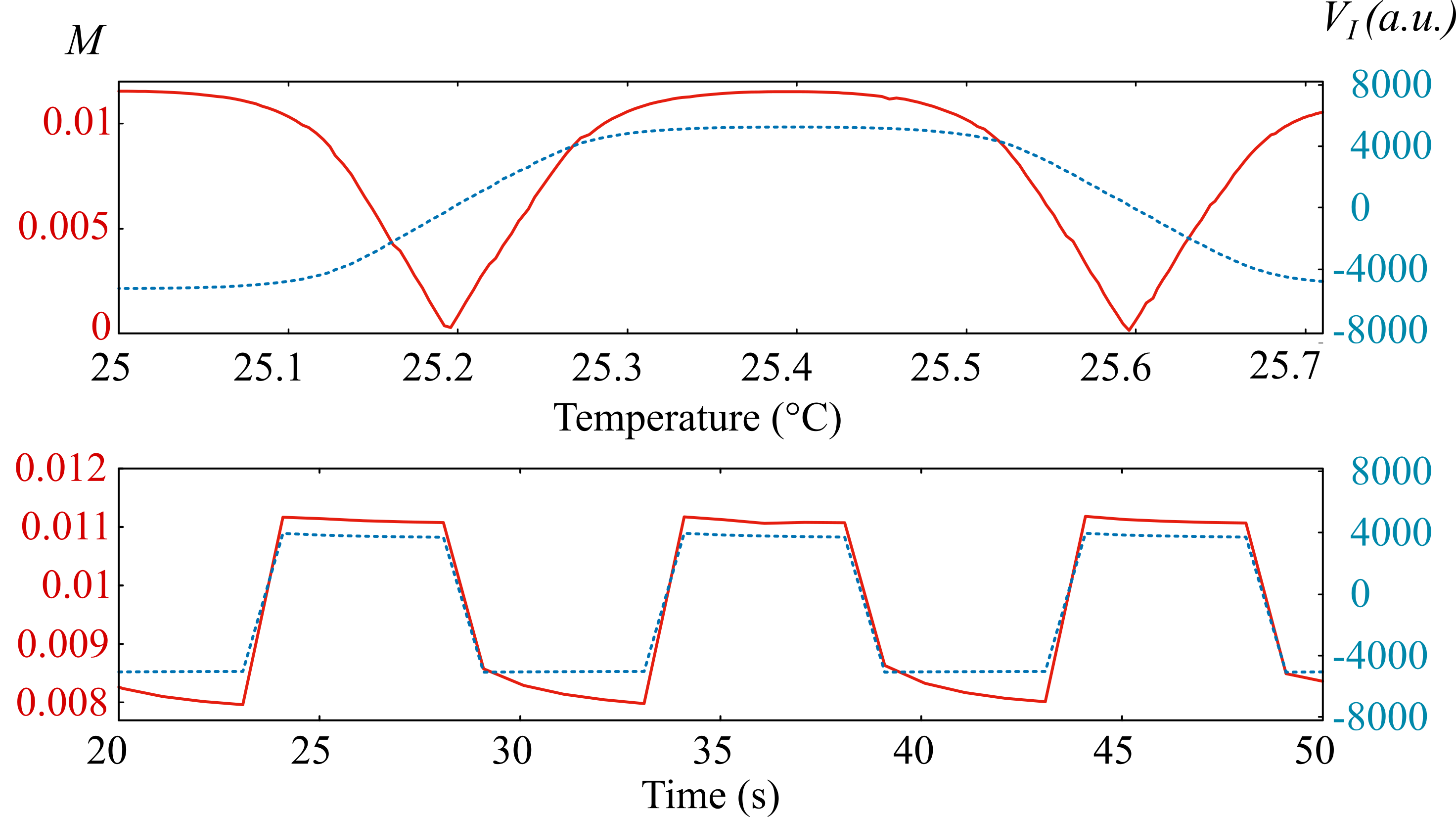}
	\end{adjustwidth}
	\caption{Top: Plots of $M$ (red) and digital error signal $V_{\textrm{I}}$ as a function of the EOM temperature.
Bottom: Plots of $M$ (red) and digital error signal $V_{\textrm{I}}$ as a function of time, during which we apply a square function of $0.1$ Hz frequency and $\pm196$ V amplitude on the DC electrodes.}
	\label{Tinversion}
\end{figure}

\indent The RAM signal $M$ and the digital error signal $V_{\textrm{I}}$ are plotted on Fig. \ref{Tinversion}. $M$ exhibits some cancellation points for particular temperatures of the crystal, at which the in-phase error signal undergoes a sign toggle. The EOM temperature lock setpoint is tuned to one of these zero-crossings before the RAM lock is engaged. The bottom graph shows the RAM and error signal behaviors when the EOM DC input is modulated with a square function of $0.1$ Hz frequency and $\pm196$ V amplitude. The dynamic range provided by the EOM DC port is sufficient to compensate RAM fluctuations of over 0.3\%, which is sufficient. We estimate that the bandwith of the EOM temperature control is limited to a few tenths of hertz by the thermal response of the EOM crystal.\\

\indent The plot of $M$ as a function of the EOM temperature shown Fig. \ref{Tinversion} exhibits RAM cancellations and error signal sign inversions every $\sim 0.4\ ^\circ$C. This value is consistent with theoretical calculations of the phase shift:
\begin{equation} \label{e2}
\Delta \phi=\frac{2 \pi \, l_0}{\lambda}\Bigl( T-T_0\Bigr) \left(\frac{\Delta n_{\textrm{e}}}{\Delta T}-\frac{\Delta n_{\textrm{o}}}{\Delta T}\right)\biggl[ 1+ \kappa \, \Bigl( T-T_0 \Bigr) \biggr]
\end{equation}

\begin{figure}[h!]
	\begin{adjustwidth}{0cm}{0cm}
		\center
		\includegraphics[width=9cm]{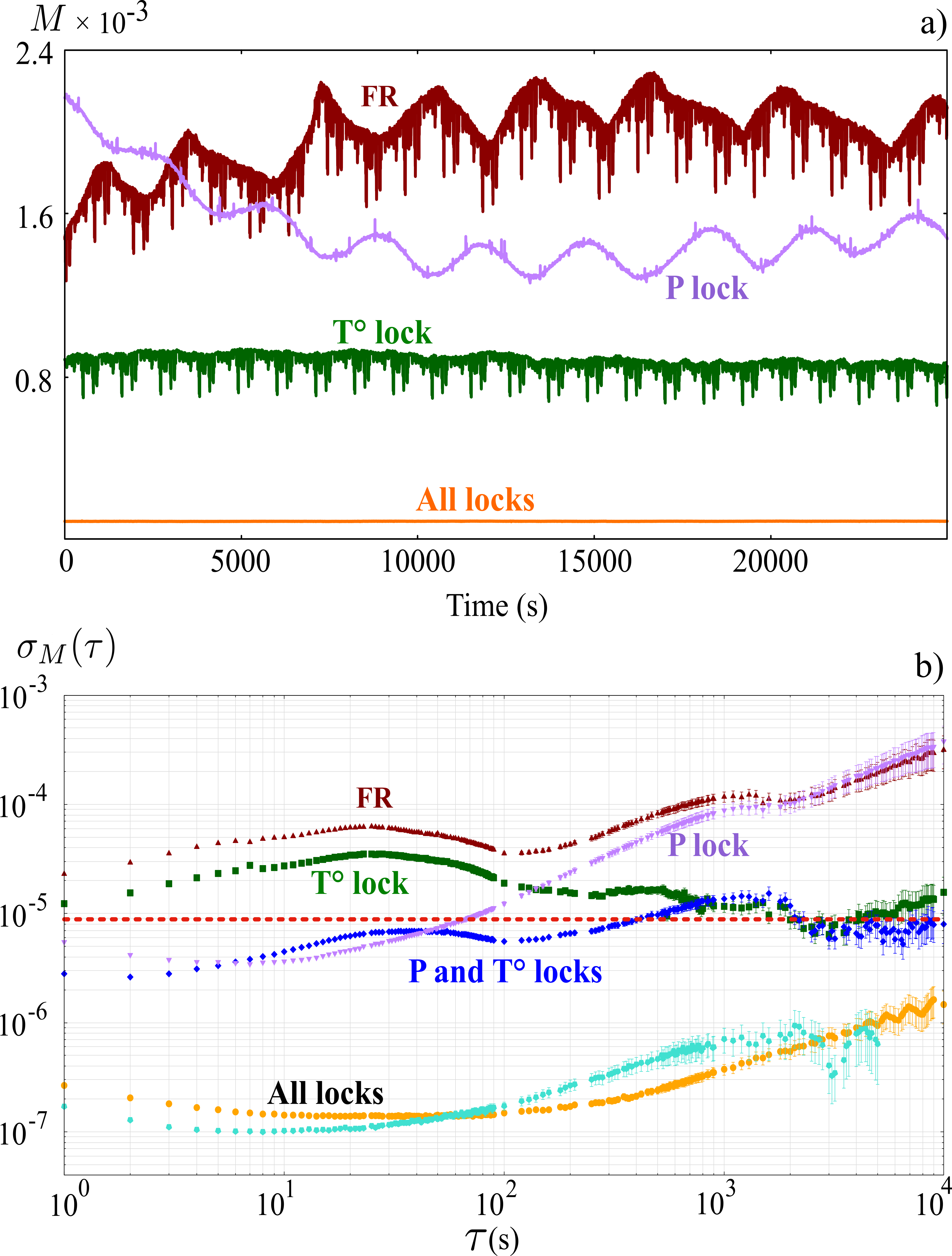}
	\end{adjustwidth}
	\caption{
	For both figures, the brown, green and pink curves are the RAM signal when the RAM servo loop is off, respectively free running (FR), with temperature stabilization (T$^{\circ}$ lock) and with power lock (P lock). The orange line (All locks) is the case when the RAM control is activated.
	a): out-of-loop signal monitored during $25000$ seconds on PD2. 
	b): Allan deviation of the out-of-loop RAM signal. Blue: both the temperature and power stabilizations are active (P and T$^{\circ}$ locks). Orange: polarizer P2 is turned at 0°. Turquoise: P2 is turned at 45°. Red dashed line: thermal noise limit.}
	\label{results}
\end{figure}

As the thermal expansion coefficient of the LiNbO$_3$ crystal for the longitudinal axis is $\kappa=1.54 \times 10^{-5}$ \cite{Jundt97}, the second term is negligible if $|T-T_0|<1$ and it is always the case in our experiment since the temperature control is $\pm 20$~mK. Only the first term is taken in account. The variations of $n_{\textrm{e}}$ and $n_{\textrm{o}}$ as a function of temperature are given in \cite{Shen92} for 1340~nm, and with these values, we obtain $\Delta \phi=2\pi$ phase shift for $|T-T_0|=0.66^\circ\textrm{C}$. Our laser emits at 1542~nm, but the Sellmeier coefficients are almost equal and we get a theoretical value close to the observation with these coefficients.\\
\indent Figure \ref{results}-a) shows the evolution of $M$ in time when all locks are off (brown curve), with temperature stabilization at $T=25.18^\circ \textrm{C}$ (green curve), with laser power stabilization (pink curve), and with all locks including RAM engaged (orange curve).  With the power lock on, sharp variations of free running RAM are erased. However, a fluctuation of a period of $2500$ s is still clearly visible on the free running data, and is due to temperature fluctuations of $\pm 1^\circ\textrm{C}$ in the laboratory. These RAM fluctuations are not fully compensated by the temperature servo of the EOM crystal and the power lock. We assume that these fluctuations emanate from the etalon effect in other optics than the EOM, because they are not temperature controlled unlike the EOM crystal.\\
\indent Fig. \ref{results}-b) shows the Allan deviation of the out-of-loop RAM index $M$. The free-running RAM (brown curve) is above $10^{-5}$ at all integration times. The green
curve is obtained when the EOM temperature lock is enabled. While the gain is marginal at short-term, the long-term drift is strongly reduced. The pink curve is obtained when the laser
power stabilization is enabled. There is a much higher gain at short term, but a strong drift after 10 s. With both temperature and laser power locked (blue curve), there is a factor of 5 to 10 reduction of RAM at all integration times. Fluctuations of $M$ are below $10^{-5}$ for $\tau$ between 1~s and 400~s, with a minimum at $3 \times 10^{-6}$. This is below the level set by the thermal noise limit of our cavity (red dashed line).\\
\indent The orange and blue curves are obtained with the RAM control enabled for two different output polarizers angles (0$^\circ$ and 45$^\circ$). There is a gain of over 100 compared to the free-running situation, and the RAM stability is well below the level set by the cavity thermal noise for integration times up to $10^4$~s. At long term ($\tau > 1000$ s), the RAM stability is still in the $5 \times 10^{-7}$ domain. By comparison with the free-running case, the long term drift of the RAM is strongly mitigated and it should not be a cause for a long-term drift of the cavity.


\section{Discussion} \label{sec:disc}

We have estimated the expected RAM contribution to the fractional frequency stability of a cryogenic silicon cavity (145~mm, $\mathcal{F}=2.5\times10^5$) stabilized laser using Eq. (\ref{sigmaram}). The RAM contribution has to be lower than the expected cavity thermal noise, which is $\sigma_{y}^{\textrm{RAM}}= 3 \times 10^{-17}$. This sets a limit $\sigma_{M}<8.8\times10^{-6}$.\\
\indent When the in-loop RAM level is minimized with $P_2$ turned at $0 ^\circ$, we reach a minimum RAM instability of $1.4 \times 10^{-7}$ for $\tau$ around 60~s, corresponding to a RAM-induced fractional frequency instability of $5 \times 10^{-19}$. We also tried to turn the polarizer of the in-loop photodiode $P_2$ at $45^\circ$, in order to increase the in-loop RAM signal and the control loop sensitivity. In this case, we achieved a slightly better result with $\sigma^{\textrm{RAM}}_y= 3.4 \times 10^{-19}$ for short integration time.\\
\indent The RAM stability shown Fig. \ref{results}-b) indicates that very low levels of RAM can be obtained by combining careful alignment of the input and output polarizers, laser power stabilization and EOM temperature stabilization. The level obtained in this configuration is competitive with both active RAM stabilization \cite{Li2012, Zhang2014} and wedged EOMs RAM levels \cite{Li2016, Tai2016} for $\tau < 10$~s. When adding the active RAM correction through the EOM DC port, our results surpass both the best active \cite{Li2014, Zhang2014} and passive \cite{Bi2019} configurations, with a RAM below $4\times10^{-7}$ from 1 to 1000~s. While the reduction of the fluctuations of the RAM is close to a factor $\sim 5$ in \cite{Zhang2014}, there is a gain over 100 compared to the free-running situation in our setup, and the RAM stability is well below the level set by the cavity thermal noise for integration times up to $10^4$~s. 
This confirms that RAM in EOMs can be pushed down to very low levels by acting passively and/or actively on critical aspects including the EOM temperature, the optics alignment and back reflections, the input and output polarization alignments, the EOM DC bias and the laser power stability.
This RAM-induced fractional frequency instability meets the requirements of our current project and is well below any current ultra-stable laser performance.\\

\section{Conclusion} 

We have achieved a reduction of a free-space EOM RAM  below $4 \times 10^{-7}$ for integration times between 1~s and 1000~s. This was made possible by the combination of laser power stabilization, EOM crystal temperature stabilization and active RAM compensation through the EOM DC port.
In addition, digital signal processing yields great flexibility and repeatability, in comparison with analog control circuits, prevents any hysteresis effects and allows to preserve performances day after day. There is room for improvement by fine-tuning the laser power control, but also by better isolating the whole optical bench from the temperature fluctuations of the laboratory.\\
\indent The RAM-induced fractional frequency instability is well below the thermal noise floor of $3 \times 10^{-17}$ computed for our single-crystal silicon cryogenic cavity. For integration times greater than $10^4$~s, the drift induced by RAM is expected to be lower than the drift of the ultra-stable cavity. These performances are then even suitable for next-generation ultra-stable cavities with enhanced stabilities.\\

\section*{Acknowledgements}
The authors would like to thank the national network for
time and frequency LabEx FIRST-TF (ANR-10-LABX-0048-01), the EIPHI Graduate School (contract ANR-17-EURE-0002), the EquipeX OSCILLATOR-IMP (ANR-11-EQPX-0033) and the Région Bourgogne Franche-Comté for supports and funding. The authors also thank
Philippe Abbé and Benoît Dubois for electronical and mechanical support.

\bibliography{biblio}

\end{document}